\documentclass[twocolumn,ol,showpacs,preprintnumbers,amsmath,amssymb]{revtex4}
\usepackage{mathbbol}
\usepackage{amssymb}
\usepackage{amsmath}
\usepackage{graphicx}
\usepackage{epsfig}
\usepackage{subfigure}
\usepackage{amsfonts}
\usepackage{CJK}

\begin{document}

\title{Experimental realization and self-testing of semisymmetric informationally complete measurements via a one-dimensional photonic quantum walk}
\author{Xu Xu$^*$}
\author{Han-Yu Cheng$\footnote{These authors contributed equally to this work}$}
\author{Meng-Yun Ma}
\author{Chao-Jie Sun}
\author{Yan Wang}
\author{Li-Jiong Shen}
\author{Zhe Sun}
\email{sunzhe@hznu.edu.cn}
\author{Qi-Ping Su}
\author{Chui-Ping Yang}
\email{yangcp@hznu.edu.cn}
\author{Yong-Nan Sun}
\email{synan@hznu.edu.cn}
\address{School of Physics, Hangzhou Normal University, Hangzhou 311121, China}


\begin{abstract}
Generalized quantum measurements play a crucial role in quantum mechanics, and symmetric informationally complete positive operator-valued measurements (SIC POVMs) provide a powerful and flexible framework for extracting information from quantum systems. However, the existence of SIC-POVMs in every finite dimension remains an open question, which has stimulated extensive research into alternative classes of POVMs. Recently, Geng $et$ $al$. [Phys. Rev. Lett. 126, 100401 (2021)] proposed a broader class of SIC POVM, called semisymmetric informationally complete POVM (semi-SIC POVM), which extends beyond SIC POVM. In this work, we focus on the four-outcome POVMs and experimentally realize the semi-SIC POVMs using a one-dimensional discrete-time quantum walk. Additionally, employing single photons and linear optics, we perform an experimental self-testing of semi-SIC POVMs in the semi-device-independent manner. Our results pave the way for exploring quantum certification with generalized quantum measurements.
\end{abstract}



\maketitle

\section{INTRODUCTION}

Quantum measurements are essential tools in quantum information processing for characterizing quantum states. Generalized quantum measurement, also known as positive operator-valued measurement (POVM), provides an effective method for extracting information that can significantly outperform von Neumann projective measurements. In particular, nonprojective POVMs are widely used in quantum theory and offer concrete advantages over projective measurements in several tasks. For example, nonprojective POVMs enhance the entanglement detection \cite{Shang2018, Bae2019} and the unambiguous state discrimination of nonorthogonal states \cite{Dieks1988, Peres1988}. They have also found applications in quantum cryptography \cite{Charles1992} and randomness generation \cite{Brask2017}.

Symmetric informationally complete POVM (SIC POVM) is an optimal type of POVM, which arises from the property that a SIC POVM is comprised of rank-one operators \cite{Renes2004}. A SIC POVM $\{E_{x}\}^{d^{2}}_{x=1}$ should satisfy three conditions: (1) $E_{x}$ is rank one for all $x \in \{1,...,d^{2}\}$, (2) the symmetry condition, Tr$[E_{x}E_{y}]=B$ for all $x\neq y$, (3) Tr$[E_{x}]=a$ for all $x \in \{1,...,d^{2}\}$ where $B$ and $a$ are constants and $d$ is the dimension of the underlying Hilbert space.

SIC POVMs lie at the intersection of mathematics and physics. In mathematics, SIC POVMs are deeply connected to prominent open problems in algebraic number theory, notably Hilbert's 12th problem \cite{Appleby2017}. In physics, they play a central role in quantum information theory and are employed in protocols such as optimal quantum state tomography \cite{Wootters1989, Scott2006,Xue2023}, dimension witness \cite{Brunner2013}, randomness generation \cite{Acin2016, Andersson2018}, quantum key distribution \cite{Fuchs2003}, and quantum certification \cite{Mironowicz2019}.

Despite the rapid growth of interest in SIC POVMs, a proof of their existence in all finite dimensions remains elusive (Zauner's conjecture) \cite{Zauner2011}. Exact and numerical solutions for SIC POVMs in various dimensions are currently being explored \cite{Appleby2019, Appleby2018, Hughston2016, Scott2010}. Meanwhile, various classes of quantum measurements, such as symmetric measurements \cite{Katarzyna2022} and equioverlapping measurements \cite{Feng2024,Guo2025}, have been extensively studied. Recently, Geng $et$ $al$. \cite{Geng2021} introduced a broader class of SIC POVMs, called semisymmetric informationally complete (semi-SIC) POVMs, which extends beyond SIC POVMs. It has been shown that, the rank-one condition of SIC POVMs is crucial, when this condition is relaxed, SIC POVMs can exist in all dimensions \cite{Spectrosc2007, Gour2014}, but they cease to be optimal. Moreover, the symmetry condition cannot be relaxed, as it constitutes the core defining characteristic of a SIC POVM. Hence, Geng $et$ $al$. showed that semi-SIC POVMs can be derived from SIC POVMs by dropping the equal-trace requirement of SIC POVMs, giving rise to several novel phenomena.

In the meantime, the increasing complexity of quantum systems has created a growing need for the certification of quantum states and measurements. Due to noise and technical imperfections inherent in quantum devices, a trend of certifying quantum states and measurements in a device-independent (DI) manner has emerged \cite{Kaniewski2016,Bancal2018,Renou2018}, which is referred to as self-testing. In the fully DI framework, the quantum device is treated as a black box, and the quantum system is characterized solely by measurement statistics, without relying on assumptions regarding the internal functioning of the devices \cite{Gomez2016,Smania2020}.

However, due to the difficulty of implementing fully DI certification, increasing attention has been directed toward the semi-device-independent (SDI) approach \cite{Pawlowski2011,Gallego2010,Bowles2013}. The primary advantage of SDI methods implies the use of prepare-and-measure (PAM) framework, which is experimentally more accessible and robust to realistic imperfections \cite{Hendrych2012,Ahrens2012,DAmbrosio2014,Ahrens2014,Sun2016,Sun2020}. Consequently, much research effort has been directed toward exploring the problem of self-testing nonprojective measurements in the SDI scenario \cite{Jonathan2021,Tavakoli2020}.

In this work, we investigate semi-SIC POVMs and experimentally realize them via a one-dimensional discrete-time quantum walk. By varying the parameter B, we implement three distinct semi-SIC POVMs and compare their characteristics with those of standard SIC-POVMs. Furthermore, we perform semi-device-independent (SDI) self-testing of these measurements by treating the state-preparation and measurement devices as black boxes and relying solely on the observed input–output statistics. This approach aims to generate correlations that, according to quantum theory, are uniquely achievable with semi-SIC POVMs. Employing single photons and linear optics, we experimentally demonstrate this SDI self-test under the bounded-dimension assumption.

\section{Theory}

\begin{figure}[h!]
\centering\includegraphics[width=8.8cm]{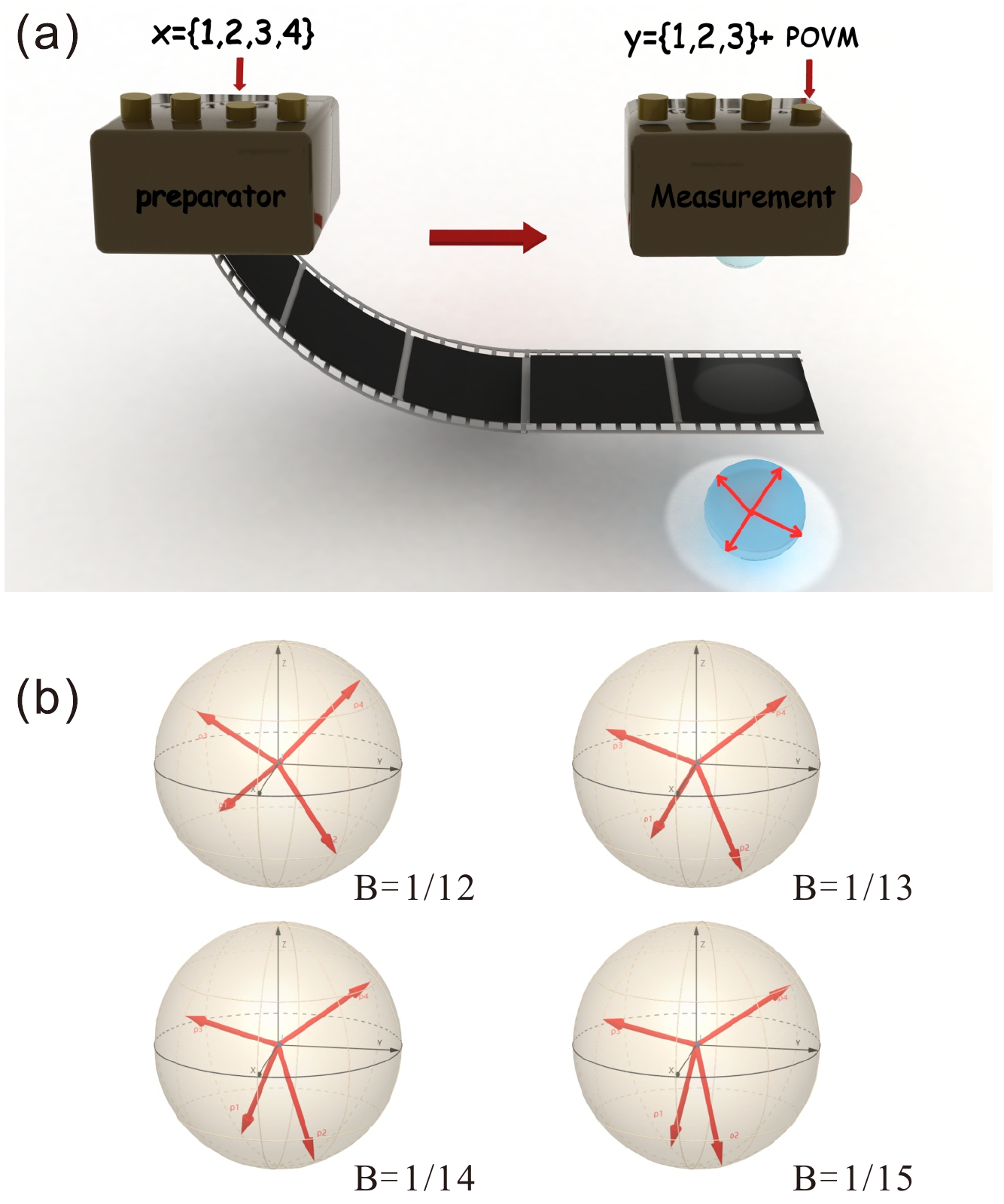}
\caption{Scenario considered for certifying the semi-SIC POVMs. (a) Alice prepares one of four different quantum states and sends to Bob, Bob performs either one of three different dichotomic measurements or a four-outcome povm measurement. (b) The Bloch vectors of semi-SIC POVMs for $B$ =1/12, 1/13, 1/14, and 1/15. When $B$ = 1/12, the semi-SIC POVM reduces to the SIC-POVM.}
\end{figure}

First, we analyze the two-dimensional semi-SIC POVMs. Compared to the conventional SIC-POVMs, the semi-SIC POVM still preserves the rank-one and symmetry conditions, but relaxes the constraint that each element has a constant trace. In other words, the semi-SIC POVMs are derived from SIC POVMs by dropping the condition Tr$[E_{x}] = 1/d$ for all outcomes $i = (1, . . . , d^{2})$. More specifically, the trace of each element of a semi-SIC POVM can take two distinct values. In the two-dimensional case, semi-SIC POVMs have the following elements \cite{Geng2021}:
\begin{equation}
\begin{aligned}
&E_{1} =a_{-}\vert\psi_{1}\rangle\langle\psi_{1}\vert,\;\;\,E_{2} =a_{-}\vert\psi_{2}\rangle\langle\psi_{2}\vert,\\
&E_{3} =a_{+} \vert\psi_{3}\rangle\langle\psi_{3}\vert,\;\;\,\,E_{4} =a_{+} \vert\psi_{4}\rangle\langle\psi_{4}\vert,\\
\end{aligned}
\end{equation}
with $a_{\pm}=(1\pm\sqrt{1-12B})/2$, and the two-dimensional vectors ${\vert\psi_{x}\rangle}$ given by
\begin{equation}
\begin{aligned}
&\vert\psi_{1}\rangle =\vert0\rangle,\quad\quad\quad\quad\quad\quad\;\;\,\,\vert\psi_{2}\rangle =r\vert0\rangle+\sqrt{1-r^{2}}\vert1\rangle,\\
&\vert\psi_{3}\rangle =\dfrac{1}{\sqrt{3}}\vert0\rangle-\sqrt{\dfrac{2}{3}}e^{i\theta}\vert1\rangle,\,\vert\psi_{4}\rangle =\dfrac{1}{\sqrt{3}}\vert0\rangle-\sqrt{\dfrac{2}{3}}e^{-i\theta}\vert1\rangle,\\
\end{aligned}
\nonumber
\end{equation}
the value of $r$ and $\theta$ can be derived from the following expression
\begin{equation}
\begin{aligned}
&r=\dfrac{2\sqrt{B}}{1-\sqrt{1-12B}},\;\,\,\theta= \rm{cos}^{-1}(\dfrac{\sqrt{1-8B-\sqrt{1-12B}}}{4\sqrt{B}}).\\
\end{aligned}
\nonumber
\end{equation}
Thus, the parameter $B$ plays a crucial role in characterizing semi-SIC POVMs: varying $B$ generates a family of distinct semi-SIC POVMs. In the two-dimensional case, $B$ is constrained to the range $B\in(1/16,1/12]$. Notably, when $B$ = 1/12, the semi-SIC POVM reduces to a standard SIC-POVM. In the subsequent experiments, we examine four representative cases: $B$ = 1/12, 1/13, 1/14, and 1/15.

\begin{figure*}
\centering \includegraphics[width=7in]{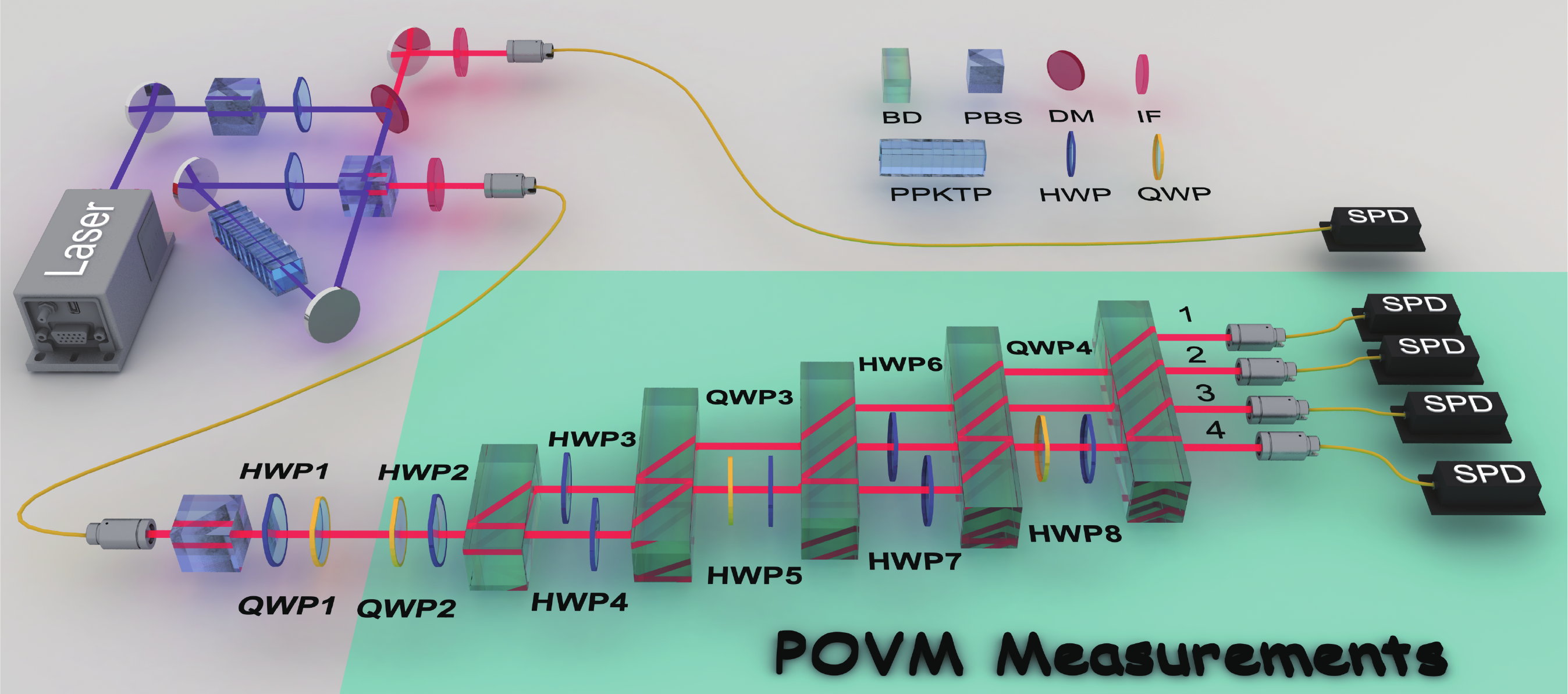}
\protect\caption{Experimental setup. Semi-SIC POVMs are implemented via a five-step quantum walk. In a one-dimensional discrete-time quantum walk, the position of the walker and the coin state are encoded by the spatial modes and the polarization states of the single photons, respectively. The HWP1 and QWP1 are used to prepare the initial state. The POVM measurements consist of five beam displacers and ten wave plates. The flipping of the site-dependent coin is enabled with different HWPs and QWPs in the optical path. PBS, polarization beam splitter; HWP, half-wave plate; QWP, quarter-wave plate; IF, interference filter; BD, beam displacer; DM, dichroic mirror.} 
\end{figure*}

Second, we investigate the self-testing of semi-SIC POVMs in the SDI manner. To self-test semi-SIC POVMs, we consider the PAM scenario illustrated in Fig. 1(a). A sender, Alice, selects one of four possible classical inputs $x \in$ $\{1,2,3,4\}$ and encodes it into a quantum state $\rho_{x}$, then sends it to Bob. Bob has two options. First, he can choose $y \in \{1,2,3\}$ and perform a dichotomic quantum measurement $M_{b|y}$ with outcome $b \in \{0,1\}$. Alternatively, Bob can select the fourth input ($y=4$) and perform a four-outcome POVM $M_{b|y=4}$ for $b \in \{1,2,3,4\}$. The correlation is described by the probability distribution $p(b|x,y)=\rm Tr(\rho_{x}M_{b|y})$. A self-test is performed when the observed statistics of measurement outcomes suffice to certify that the prepared quantum states and measurement operators correspond to their claimed descriptions. Within the PM scenario, a self-test of semi-SIC POVMs can be achieved using a linear witness $W$, which is expressed as \cite{Drotos2024,Tavakoli2020}
\begin{equation}
W=W_{1}-k \sum_{x=1}^{4}P(b=x|x,y=4),
\end{equation}
where $k$ is positive constant and $W_{1}$ is a witness obtained through the dichotomic quantum measurements, namely $W_{1}=\sum_{x,y}\omega_{xy}E_{xy}$. $E_{xy}$ is the expectation value of an observable and $\omega_{xy}$ represents the witness matrices \cite{Drotos2024}. Our objective is to construct a witness $W$ such that its maximum value $Q$ attainable with qubit systems self-tests the target semi-SIC POVMs. According to the work \cite{Drotos2024}, the maximum value $Q$ of distinct semi-SIC POVMs is 
\begin{equation}
Q=24\sqrt{\dfrac{B}{24B-1}}.
\end{equation}
We measure the value of the linear witness $W$ according to Eq. (2). Once the measured value of $W$ approaches the theoretical maximum value $Q$, the semi-SIC POVMs can be self-tested. Finally, we perform self-testing of semi-SIC POVMs for $B$ = 1/13, 1/14, and 1/15. The corresponding Bloch vectors, shown in Fig. 1(b), are equivalent to the semi-SIC POVMs defined in Eq. (1) up to a unitary or antiunitary transformation \cite{Drotos2024}.

\section{Experiment}
 
The experimental setup is illustrated in Fig.~2. A polarization beam splitter (PBS) and a half-wave plate (HWP) are employed to control the polarization of the pump beam. Inside the Sagnac interferometer, a 405 nm laser pumps a periodically poled KTiOPO4 (PPKTP) nonlinear crystal to generate photon pairs at 810 nm. After passing through an interference filter, the photons are separately coupled into single-mode fibers. The detection of one photon serves as a trigger to herald the presence of the other, thereby preparing a single-photon state. Photons are detected by avalanche photodiodes within a 3 ns coincidence window. At a pump power of 2.67 mW, the total coincidence count is about $3 \times 10^{4}$ per second.

In our experiment, quantum state is encoded in the horizontal ($\vert H \rangle=\vert 0 \rangle$) and vertical ($\vert V \rangle=\vert 1 \rangle$) polarization of photon. After passing through two wave plates (HWP1 and QWP1), the single photon state is prepared in the desired initial state $|\psi\rangle=\operatorname{cos}\frac{\theta}{2}|H\rangle+ e^{i\phi}\operatorname{sin}\frac{\theta}{2}|V\rangle $. The four-outcome POVMs are realized through a suitably configured discrete-time quantum walk \cite{Kurzynski2013,Bian2015,yuan2015,Fan2022}. Consequently, the desired single-qubit semi-SIC POVM is implemented using a five-step quantum walk. In general, the evolution operator ($U$) of a quantum walk consists of a conditional translation operator ($T$) and a coin operator ($C$), defined as $U = TC$. The site-dependent coin operations $C_{x,n}$ for the $n$th step and the conditional translation operator $T$ are \cite{Kurzynski2013}:
\begin{equation}
\begin{aligned}
&C_{x,n}=\left(\begin{array}{cc}\cos\theta_{x}&e^{-i\beta}\sin\theta_{x}\\\\-e^{i\beta}\sin\theta_{x}&\cos\theta_{x}\end{array}\right), \\
&T=\sum_x|x+1,H\rangle\langle x,H|+|x-1,V\rangle\langle x,V|.
\end{aligned}
\end{equation}

The walker's position ($x$) and the coin state are encoded in the spatial mode and the polarization state of the single photon, respectively. Site-dependent coin operations $C_{x,n}$ are implemented using wave plates positioned in specific spatial paths, while the conditional translation operator $T$ is realized through a cascaded interferometric network. The quantum measurements are performed via a quantum walk consisting of five beam displacers (BDs) and ten wave plates (QWP2, HWP2, HWP3, HWP4, QWP3, HWP5, HWP6, HWP7, QWP4, and HWP8). Sequential pairs of BDs form interferometers, which require mutual alignment of their optical axes. In the experiment, the interferometric extinction ratio remains stable at 220:1 over a period of approximately one hour. There are four output ports in the experimental setup, each corresponding to an outcome of a semi-SIC POVM element ($E_{i}$). For a properly engineered quantum walk procedure, the walker initialized in different coin states evolves to distinct position distributions. By performing position measurements on the walker in the final step, one can implement the semi-SIC POVM.

\subsection{Experimental realization of semi-SIC POVMs}

To implement semi-SIC POVMs using a five-step quantum walk, the initial state $\left|\psi\right\rangle=\frac1{\sqrt{2}}\left(\left|H\right\rangle+\left|V\right\rangle\right)$ is prepared by adjusting wave plates HWP1 and QWP1. The conditional translation operator $T$ is realized through a cascaded interferometric network. Crucially, the successful implementation of the semi-SIC POVM relies on the experimental realization of site-dependent coin operations $C_{x,n}$. For example, in order to realize the semi-SIC POVM with $B$ = 1/13, the five-step of the coin operator are 
\begin{equation}
\begin{aligned}
&C_{0,1}=\rm I,\quad
C_{1,2}=\left(\begin{array}{cc}0.6011&0.7992\\\\0.7992&-0.6011\end{array}\right), \\
&C_{-1,2}=C_{-1,4}=\sigma_x, \\
&C_{0,3}=\left(\begin{array}{cc}0.5551&0.8318 \\\\0.8318 &-0.5551\end{array}\right),\\
&C_{1,4}=\left(\begin{array}{cc}0.6941&0.7199\\\\0.7199&-0.6941\end{array}\right),\\
&C_{0,5}=\left(\begin{array}{cc}-0.0771+0.7029i&0.7028+0.0771i\\\\0.0771+0.7028i&-0.7028+0.0771i\end{array}\right).\\
\end{aligned}
\end{equation}
When $B$ takes other values, we can calculate site-dependent coin operations through Eq. (1) and Eq. (4), thereby enabling us to implement arbitrary semi-SIC POVMs. Subsequently, the angles of the HWP and QWP are specifically set in order to realize the site-dependent coin rotations $C_{x,n}$, as shown in Table \uppercase\expandafter{\romannumeral1}. After performing position measurements on the walker during the final step, we can derive the probability distributions $P$ for the four output ports (1, 2, 3, 4), the probability of measuring the $i$th semi-SIC POVM element on a state $\rho$ is given by $P_i=\rm Tr[\rho E_{i}]$ ($i = (1,2,3,4)$).

\begin{table}[h!]
\centering
\caption{The angles of HWPs and QWPs used to realize semi-SIC POVMs for $B$= 1/12, 1/13, 1/14, and 1/15, while the angle of HWP4 and HWP7 are fixed at 45$^\circ$.}
\begin{tabular}{c@{\hspace{0.6cm}}c@{\hspace{0.6cm}}c@{\hspace{0.6cm}}c@{\hspace{0.6cm}}c@{\hspace{0.6cm}}c}
  \hline\hline
  B & HWP3 & HWP5 & HWP6 & QWP4 & HWP8 \\
 
  $1/12$ & 22.5$^\circ$ & 22.5$^\circ$ & 17.63$^\circ$ & 60.34$^\circ$ & 7.67$^\circ$   \\
 
  $1/13$ & 26.53$^\circ$ & 28.14$^\circ$ & 23.02$^\circ$ & 96.26$^\circ$ & 25.63$^\circ$   \\
 
  $1/14$ & 28.05$^\circ$ & 31.86$^\circ$ & 24.96$^\circ$ & 67.03$^\circ$ & 11.01$^\circ$   \\
 
  $1/15$ & 29.14$^\circ$ & 36$^\circ$ & 26.31$^\circ$ & 22.19$^\circ$ & -11.4$^\circ$   \\
  \hline\hline
\end{tabular}
\end{table}

Figure 3 illustrates the probability distributions of the five-step quantum walk for semi-SIC POVMs, derived by estimating the frequency of repeated measurements, showing a high degree of agreement with theoretical predictions. For example, consider the experimental distribution of the quantum walk with $B = 1/13$, where the probabilities after five steps are $P(E_{1}) = 0.1832 \pm 0.0023$, $P(E_{2}) = 0.3563 \pm 0.0021$, $P(E_{3}) = 0.2316 \pm 0.0022$, and $P(E_{4}) = 0.2289 \pm 0.0015$, while the corresponding theoretical results are $P(E_{1}) = 0.1807$, $P(E_{2}) = 0.3584$, $P(E_{3}) = 0.2305$, and $P(E_{4}) = 0.2305$. These experimental results closely match the theoretical predictions with a maximum deviation of less than 0.005, thereby confirming the capability of semi-SIC POVMs to measure arbitrary quantum states. For comparison, we also performed measurements based on the SIC POVM. We note that when $B = 1/12$, the semi-SIC POVM reduces to the SIC POVM; the measured probability distribution for this case is shown in Fig.~3(a).

\begin{figure}[h!]
\centering
\includegraphics[width=3.5in]{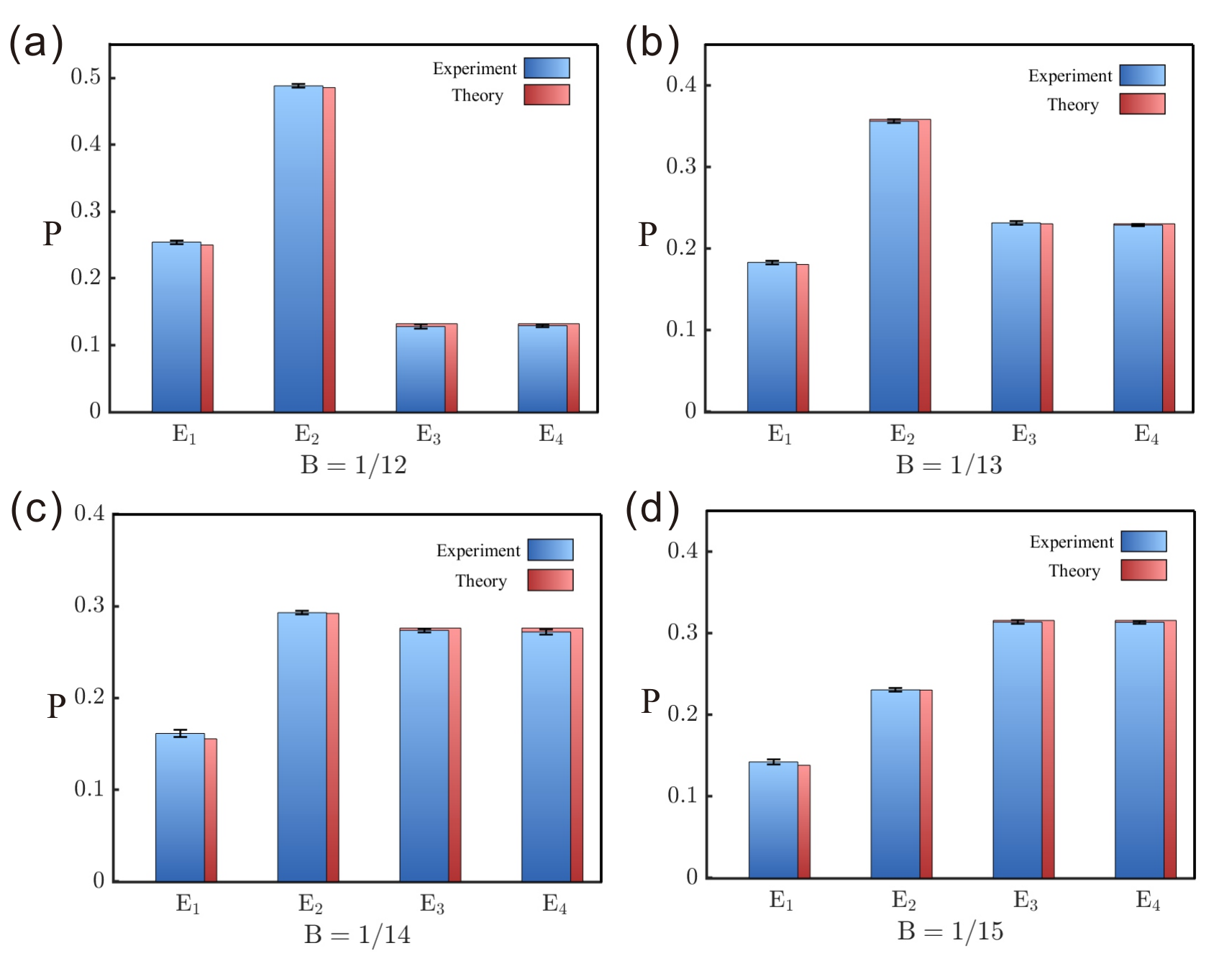}
\caption{Experimental results of semi-SIC POVMs via a photonic quantum walk. The measured probability distributions for semi-SIC POVMs with $B = 1/12$, $1/13$, $1/14$, and $1/15$ show close agreement with theoretical predictions, with a maximum deviation of less than $0.005$.}
\end{figure}

\subsection{Self-testing of semi-SIC POVMs}
The second experiment aims to demonstrate the self-testing of semi-SIC POVMs. As shown in Fig.~1(a), to realize the self-testing of semi-SIC POVMs, Alice prepares four initial states, after which Bob randomly measures the received states and substitutes the results into Eq.~(2). By comparing the linear witness $W$ from Eq.~(2) with the maximum value $Q$, we can determine whether Bob's measurement corresponds to a semi-SIC POVM. In the experiment, we realize the self-testing for different types of semi-SIC POVMs, specifically when $B$ = 1/13, 1/14, and 1/15, we perform the experiment individually. Taking $B$ = 1/13 as an example, we describe the entire experimental process in detail. For $B$ = 1/13, the four initial states are
\begin{equation}
\begin{aligned}
&\left|\psi_{1}\right\rangle  =0.3409|H\rangle-(0.6648+0.6648i)|V\rangle,   \\
&\left|\psi_{2}\right\rangle  =0.3409|H\rangle+(0.6648+0.6648i)|V\rangle,   \\
&\left|\psi_{3}\right\rangle  =0.8468|H\rangle+(0.3761-0.3761i)|V\rangle,   \\
&\left|\psi_{4}\right\rangle  =0.8468|H\rangle-(0.3761-0.3761i)|V\rangle,
\end{aligned}
\end{equation}
other initial states can be obtained through Eq. (1).

\begin{figure}[tbph]
\centering
\includegraphics[width=3.2in]{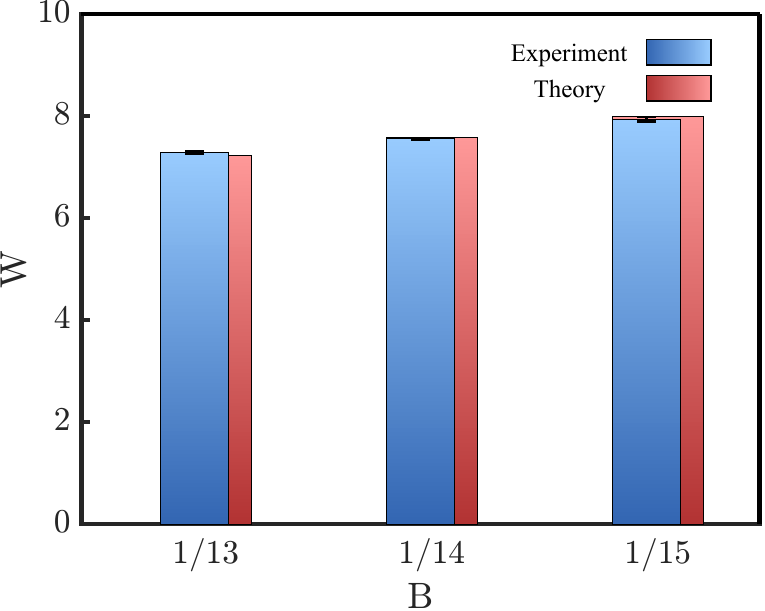}
\caption{Experimental results for self-testing of semi-SIC POVMs. The witness values $W$ for $B$ = 1/13, 1/14, and 1/15 are $7.2908\pm0.0304$, $7.5615\pm0.0149$ and $7.9393\pm0.0434$, respectively. These experimental results are close to the theoretical predictions and the error bars indicate the statistical uncertainty.}
\end{figure}

\begin{table}[h!]
\centering
\caption{The angles of HWPs and QWPs used to realize the self-testing of semi-SIC POVMs for B= 1/13, 1/14, and 1/15, the angle of HWP4 and HWP7 are fixed at 45$^\circ$.}
\begin{tabular}{c@{\hspace{1.15cm}}c@{\hspace{1.15cm}}c@{\hspace{1.15cm}}c}
  \hline\hline
  Wave plates & B=1/13 & B=1/14 & B=1/15 \\

  QWP2 & 15.28$^\circ$ & 11.41$^\circ$ & 7.56$^\circ$  \\

  HWP2 & 0.9$^\circ$ & 0.41$^\circ$ & 0.13$^\circ$  \\

  HWP3 & 26.53$^\circ$ & 28.05$^\circ$ & 29.14$^\circ$  \\

  QWP3 & -62.16$^\circ$ & -69.79$^\circ$ & -76.81$^\circ$  \\

  HWP5 & -19.31$^\circ$ & -25.63$^\circ$ & -32.05$^\circ$  \\

  HWP6 & 23.02$^\circ$ & 24.96$^\circ$ & 26.31$^\circ$  \\
  
  QWP4 & -45$^\circ$ & -45$^\circ$ & -45$^\circ$  \\

  HWP8 & -15.86$^\circ$ & -13.68$^\circ$ & -12.28$^\circ$  \\
  \hline\hline
\end{tabular}
\end{table}

We implement self-testing via a quantum walk comprising five beam displacers and ten wave plates. To achieve this, the angles of the HWPs and QWPs are configured to generate the site-dependent coin operations $C_{x,n}$, as detailed in Table \uppercase\expandafter{\romannumeral2}. For the specific case of $B = 1/13$, the coin operators for the five steps are
\begin{equation}
\begin{aligned}
&C_{0,1}=\left(\begin{array}{cc}0.9381 + 0.0614i&0.2563 - 0.2248i\\\\-0.2248 + 0.2563i&-0.0614 - 0.9381i\end{array}\right),\\
&C_{1,2}=\left(\begin{array}{cc}0.6011&0.7992\\\\0.7992&-0.6011\end{array}\right), \\
&C_{-1,2}=C_{-1,4}=\sigma_x, \\
&C_{0,3}=\left(\begin{array}{cc}-0.4281 - 0.3533i&0.8106 - 0.1866i\\\\-0.1866 + 0.8106i&0.3533 + 0.4281i\end{array}\right),\\
&C_{1,4}=\left(\begin{array}{cc}0.6941&0.7199\\\\0.7199&-0.6941\end{array}\right),\\
&C_{0,5}=\left(\begin{array}{cc}0.6882 + 0.1625i&-0.6882 + 0.1625i\\\\0.1625 - 0.6882i&-0.1625 -0.6882i\end{array}\right).\\
\end{aligned}
\end{equation}

By adjusting the angles of the HWPs and QWPs in the optical path, all input photons are directed to the four output ports (1, 2, 3, 4). In this configuration, HWP4 and HWP7 are fixed at $45^\circ$, and the coupling efficiency of each port is calibrated to ensure that the difference in detection efficiency does not exceed $5\%$. Figure 4 presents the experimental results of the witness $W$ for qubit semi-SIC POVMs with $B = 1/13$, $1/14$, and $1/15$. The corresponding witness values are $7.2908 \pm 0.0304$, $7.5615 \pm 0.0149$, and $7.9393 \pm 0.0434$, showing close agreement with the theoretical predictions of 7.2363, 7.5895, and 8, respectively. These results confirm that the device successfully implements semi-SIC POVMs, with the primary sources of discrepancy attributed to imperfections in the wave plates and photon counting statistics.

\section{Conclusion}
In this work, we investigate semi-SIC POVMs and experimentally implement them via a one-dimensional discrete-time quantum walk. In our setup, arbitrary single-qubit POVMs are realized via the discrete-time quantum walk, with quantum states encoded in the polarization degrees of freedom of photons with high fidelity. By varying the parameter $B$, we realize three distinct semi-SIC POVMs and compare their characteristics with those of standard SIC-POVMs. Furthermore, we investigate the SDI self-testing of four-outcome semi-SIC POVMs, establishing a method for certifying these quantum systems. Using single photons and linear optics, we experimentally demonstrate this SDI self-testing under a bounded-dimension assumption.

\section*{ACKNOWLEDGMENTS}
This work was supported by the National Key Research and Development Program of China (Grant No. 2024YFA1408900), the Hangzhou Joint Fund of the Zhejiang Provincial Natural Science Foundation of China under Grant No. LHZSD24A050001, the National Natural Science Foundation of China (62105086, U21A20436), Innovation Program for Quantum Science and Technology (2021ZD0301705), Hangzhou Leading Youth Innovation and Entrepreneurship Team project under Grant No. TD2024005, and Scientific Research Foundation for Scholars of HZNU (4085C50221204030).


\begin{thebibliography}{99}




\bibitem{Shang2018} Jiangwei Shang, Ali Asadian, Huangjun Zhu, and Otfried Gühne, Enhanced entanglement criterion via symmetric informationally complete measurements, Phys. Rev. A {\bf 98}, 022309 (2018).

\bibitem{Bae2019} Joonwoo Bae, Beatrix C Hiesmayr, and Daniel McNulty, Linking entanglement detection and state tomography via quantum 2-designs, New Journal of Physics {\bf 21}, 013012 (2019).

\bibitem{Dieks1988} Dennis Dieks, Overlap and distinguishability of quantum states, Physics Letters A {\bf 126}, 303-306 (1988).

\bibitem{Peres1988} Asher Peres, How to differentiate between non-orthogonal states, Physics Letters A {\bf 128}, 19 (1988).

\bibitem{Charles1992} Charles H Bennett, Quantum cryptography using any two nonorthogonal states, Phys. Rev. Lett. {\bf 68}, 3121 (1992).

\bibitem{Brask2017} J. B. Brask, A. Martin, W. Esposito, R. Houlmann, J. Bowles, H. Zbinden, and N. Brunner, Megahertz-rate semi-device-independent quantum random number generators based on unambiguous state discrimination, Phys. Rev. Appl. {\bf 7}, 054018 (2017).


\bibitem{Renes2004} J. M. Renes, R. Blume-Kohout, A. J. Scott, and C. M. Caves, J. Math. Phys. (N.Y.) {\bf 45}, 2171 (2004).







\bibitem{Appleby2017} Marcus Appleby, Steven Flammia, Gary McConnell, and Jon Yard, SICs and algebraic number theory, Foundations of Physics {\bf 47}, 1042-1059 (2017).





\bibitem{Wootters1989} William K Wootters and Brian D Fields, Optimal state-determination by mutually unbiased measurements, Annals of Physics {\bf 191}, 363-381 (1989).


\bibitem{Scott2006} Andrew J Scott, Tight informationally complete quantum measurements, Journal of Physics A: Mathematical and General {\bf 39}, 13507 (2006).



\bibitem{Xue2023} Xiaowei Wang, Xiang Zhan, Yulin Li, Lei Xiao, Gaoyan Zhu, Dengke Qu, Quan Lin, Yue Yu, and Peng Xue, Generalized Quantum Measurements on a Higher-Dimensional System via Quantum Walks, Phys. Rev. Lett. {\bf 131}, 150803 (2023).





\bibitem{Fuchs2003} Christopher A Fuchs and Masahide Sasaki, Squeezing quantum information through a classical channel: measuring the quantumness of a set of quantum states, Quantum Information Computation {\bf 3}, 377-404 (2003). 

\bibitem{Brunner2013} Nicolas Brunner, Miguel Navascues, and Tamas Vertesi, Dimension witnesses and quantum state discrimination, Phys. Rev. Lett. {\bf 110}, 150501 (2013).





\bibitem{Acin2016} Antonio Acin, Stefano Pironio, Tamas Vertesi, and Peter Wittek, Optimal randomness certification from one entangled bit, Phys. Rev. A {\bf 93}, 040102 (2016).

\bibitem{Andersson2018} Ole Andersson, Piotr Badziag, Irina Dumitru, and Adan Cabello, Device-independent certification of two bits of randomness from one entangled bit and Gisin's elegant Bell inequality, Phys. Rev. A {\bf 97}, 012314 (2018).



\bibitem{Mironowicz2019} Piotr Mironowicz and Marcin Pawlowski, Experimentally feasible semi-device-independent certification of four-outcome positive-operator-valued measurements, Phys. Rev. A {\bf 100}, 030301 (2019).




\bibitem{Zauner2011} G. Zauner, Quantum Designs: Foundations of a Noncommutative Design Theory, Int. J. Quantum. Inform. {\bf 09}, 445 (2011).


\bibitem{Appleby2019} M. Appleby and I. Bengtsson, Simplified exact SICs, J. Math. Phys. (N.Y.) {\bf 60}, 062203 (2019).


\bibitem{Appleby2018} M. Appleby, T.-Y. Chien, S. Flammia, and S. Waldron, Constructing exact symmetric informationally complete measurements from numerical solutions, J. Phys. A {\bf 51}, 165302 (2018).


\bibitem{Hughston2016} L. P. Hughston and S. M. Salamon, Surveying points in the complex projective plane, Adv. Math. {\bf 286}, 1017 (2016).


\bibitem{Scott2010} A. J. Scott and M. Grassl, Symmetric informationally complete positive-operator-valued measures: A new computer study, J. Math. Phys. (N.Y.) {\bf 51}, 042203 (2010).



\bibitem{Katarzyna2022} Katarzyna Siudzi$\rm\acute{n}$ska, All classes of informationally complete symmetric measurements in finite dimensions, Phys. Rev. A {\bf 105}, 042209 (2022).



\bibitem{Feng2024} Lingxuan Feng, Shunlong Luo, Yan Zhao, and Zhihua Guo, Equioverlapping measurements as extensions of symmetric informationally complete positive operator valued measures, Phys. Rev. A {\bf 109}, 012218 (2024).


\bibitem{Guo2025} Zhihua Guo, Yan Liu, Tsung-Lin Lee, Shunlong Luo, Equioverlapping measurements as extensions of symmetric informationally complete positive operator valued measures, Phys. Rev. A {\bf 111}, 012430 (2025).









\bibitem{Geng2021} Isabelle Jianing Geng, Kimberly Golubeva, and Gilad Gour, What are the minimal conditions required to define a symmetric informationally complete generalized measurement? Phys. Rev. Lett. {\bf 126}, 100401 (2021).






\bibitem{Spectrosc2007} D. M. Appleby, Symmetric informationally complete measurements of arbitrary rank, Opt. Spectrosc. {\bf 103}, 416 (2007).

\bibitem{Gour2014} G. Gour and A. Kalev, Construction of all general symmetric informationally complete measurements, J. Phys. A {\bf 47}, 335302 (2014).







\bibitem{Kaniewski2016} J. Kaniewski, Analytic and Nearly Optimal Self-Testing Bounds for the Clauser-Horne-Shimony-Holt and Mermin Inequalities, Phys. Rev. Lett. {\bf 111}, 070402 (2016).


\bibitem{Bancal2018} J. D. Bancal, N. Sangouard, and P. Sekatski, Noise Resistant Device-Independent Certification of Bell State Measurements, Phys. Rev. Lett. {\bf 121}, 250506 (2018).



\bibitem{Renou2018} M. O. Renou, J. Kaniewski, and N. Brunner, Self-Testing Entangled Measurements in Quantum Networks, Phys. Rev. Lett. {\bf 121}, 250507 (2018).




\bibitem{Gomez2016} E. S. G{\'o}mez, S. G{\'o}mez, P. Gonz{\'a}lez, G. Canas, J. F. Barra, A. Delgado, G. B. Xavier, A. Cabello, M. Kleinmann, T. V{\'e}rtesi, and G. Lima, Device-Independent Certification of a Nonprojective Qubit Measurement, Phys. Rev. Lett. {\bf 117}, 260401 (2016).




\bibitem{Smania2020} M. Smania, P. Mironowicz, M. Nawareg, M. Paw{\l}owski, A. Cabello, and M. Bourennane, Experimental certification of an informationally complete quantum measurement in a device-independent protocol, Optica {\bf 7}, 123 (2020).




\bibitem{Pawlowski2011} M. Paw{\l}owski, and N. Brunner, Semi-device-independent security of one-way quantum key distribution. Phys. Rev. A {\bf 84}, 010302(R) (2011).



\bibitem{Gallego2010} R. Gallego, N. Brunner, C. Hadley, and A. Ac{\'i}n, Device-Independent Tests of Classical and Quantum Dimensions, Phys. Rev. Lett. {\bf 105}, 230501 (2010).




\bibitem{Bowles2013} J. Bowles, M. T. Quintino, and N. Brunner, Certifying the Dimension of Classical and Quantum Systems in a Prepare-and-Measure Scenario with Independent Devices, Phys. Rev. Lett. {\bf 112}, 140407 (2013).





\bibitem{Hendrych2012} M. Hendrych, R. Gallego, M. Miuda, N. Brunner, A. Acin, and J. P. Torres, Experimental estimation of the dimension of classical and quantum systems, Nat. Phys. {\bf 8}, 588 (2012).



\bibitem{Ahrens2012} J. Ahrens, P. Badziag, A. Cabello, and M. Bourennane, Experimental device-independent tests of classical and quantum dimensions, Nat. Phys. {\bf 8}, 592 (2012).






\bibitem{DAmbrosio2014} V. D'Ambrosio, F. Bisesto, F. Sciarrino, J. F. Barra, G. Lima, and A. Cabello, Device-Independent Certification of High-Dimensional Quantum Systems, Phys. Rev. Lett. {\bf 112}, 140503 (2014).




\bibitem{Ahrens2014} J. Ahrens, P. Badziag, M. Paw{\l}owski, M. Zukowski, and M. Bourennane, Experimental Tests of Classical and Quantum Dimensionality, Phys. Rev. Lett. {\bf 112}, 140401 (2014).




\bibitem{Sun2016} Y.-N. Sun, Z.-D. Liu, J. Sun, G. Chen, X.-Y. Xu, Y.-C. Wu, J.-S. Tang, Y.-J. Han, C.-F. Li, and G.-C. Guo, Experimental realization of dimension witnesses based on quantum state discrimination, Phys. Rev. A {\bf 94}, 052313 (2016).






\bibitem{Sun2020} Y.-N. Sun, Z.-D. Liu, J. Bowles, G. Chen, X.-Y. Xu, J.-S. Tang, C.-F. Li, and G.-C. Guo, Experimental certification of quantum dimensions and irreducible high-dimensional quantum systems with independent devices, Optica {\bf 7}, 1073 (2020).





\bibitem{Jonathan2021} J. Steinberg, H. C. Nguyen, and M. Kleinmann, Minimal scheme for certifying three-outcome qubit measurements in the prepare-and-measure scenario, Phys. Rev. A {\bf 104}, 062431 (2021).




\bibitem{Tavakoli2020} Armin Tavakoli, Massimiliano Smania, Tamas Vertesi, Nicolas Brunner, and Mohamed Bourennane, Self-testing nonprojective quantum measurements in prepare-andmeasure experiments, Science Advances {\bf 6}, eaaw6664 (2020).


\bibitem{Drotos2024} G$\rm\acute{a}$bor Dr$\rm\acute{o}$tos, K$\rm\acute{a}$roly F. P$\rm\acute{a}$l, and Tam$\rm\acute{a}$s V$\rm\acute{e}$rtesi, Self-testing of semisymmetric informationally complete measurements in a qubit prepare-and-measure scenario, Phys. Rev. A {\bf 110}, 032427 (2024).



\bibitem{Kurzynski2013} Pawe{\l} Kurzy{\'n}ski, and Antoni W{\'o}jcik, Quantum Walk as a Generalized Measuring Device, Phys. Rev. Lett. {\bf 110}, 200404 (2013).




\bibitem{Bian2015} Zhihao Bian, Jian Li, Hao Qin, Xiang Zhan, Rong Zhang, Barry C. Sanders, and Peng Xue, Realization of Single-Qubit Positive-Operator-Valued Measurement via a One-Dimensional Photonic Quantum Walk, Phys. Rev. Lett. {\bf 114}, 203602 (2015).


\bibitem{yuan2015} Yuan-yuan Zhao, Neng-kun Yu, Pawe{\l} Kurzy{\'n}ski, Guo-yong Xiang, Chuan-Feng Li, and Guang-Can Guo, Experimental realization of generalized qubit measurements based on quantum walks, Phys. Rev. A {\bf 91}, 042101 (2015).



\bibitem{Fan2022} Qin Fan, Meng-Yun Ma, Yong-Nan Sun, Qi-Ping Su, and Chui-Ping Yang, Experimental certification of nonprojective quantum measurements under a minimum overlap assumption, Opt. Express {\bf 30}, 34441 (2022).












\end{thebibliography}
\end{document}